\documentclass[prb,twocolumn,showpacs,byrevtex,floatfix]{revtex4}

\usepackage{graphics}
\usepackage{pslatex}
\usepackage{amsmath}

\newcommand{\ie}{{\it i.e.}}
\newcommand{\eg}{{\it e.g.}}

\newcommand{\etc}{{\it etc.}}

\newcommand{\etal}{{\it et~al.}}

\newcommand{\YBCO}{YBa$_2$Cu$_3$O$_7$}

\begin{document}

\title{
  Semi-fluxons in long Josephson $0$-$\pi$-junctions.
}

\author{E.~Goldobin}
\email{gold@uni-tuebingen.de}
\homepage{http://www.geocities.com/e_goldobin}
\author{D.~Koelle}
\author{R.~Kleiner}
\affiliation{
  Physikalisches Institut - Experimentalphysik II,
  Universit\"at T\"ubingen,
  Auf der Morgenstelle 14,
  D-72076 T\"ubingen, Germany
}


\begin{abstract}
  We investigate analytically long Josephson junctions with phase $\pi$-discontinuity points. Such junctions are usually fabricated as a ramp between a superconductor like {\YBCO} with $d$-wave symmetry of the order parameter and an $s$-wave superconductor like Nb. From the top, they look like zigzags with $\pi$-jumps of the Josephson phase at the corners. These $\pi$-jumps, at certain conditions, lead to the formation of half-integer flux quanta, which we call semi-fluxons, pinned at the corners. We derive a version of sine-Gordon equation which describes the dynamics of the Josephson phase in such structures, and obtain an explicit formula which describes the shape of a semi-fluxon. Some properties of semi-fluxons are discussed. We propose a way to construct artificial $\pi$-junctions using only $s$-wave superconductors.
\end{abstract}

\maketitle

\section{Introduction}
\label{Sec:Intro}

Recent experiments with {\YBCO}-Nb ramp long Josephson junctions (LJJ) fabricated in a zigzag geometry (if viewed from the top) clearly demonstrated that due to the specific order parameter symmetry the LJJ consists of alternating facets of $0$, $\pi$, $0$, $\pi\ldots$ junctions\cite{Smilde:ZigzagPRL}. As a result, half-integer flux quanta can be spontaneously generated and trapped at the corners of the zigzag, because these are exactly the points where the order parameter of high-$T_c$ superconductor changes its sign due to a $90^\circ$ change in the direction of the Josephson contact (direction of Josephson tunneling current)\cite{Hilgenkamp-Tsuei}. Half-integer flux quanta, further called semi-fluxons (SFs), were also experimentally observed in tri-crystal grain boundary (GB) LJJs
\cite{Kirtley:SF:HTSGB,Kirtley:SF:T-dep,Sugimoto:TriCrystal:SF}. 
The presence of alternating $0$- and $\pi$-facets results in a set of $\pi$-discontinuities of the Josephson phase at the corners where $0$- and $\pi$-facets join. The possibility to fabricate such LJJs opens new perspectives for Josephson electronics (digital circuits, fluxon devices, quantum bits, \etc)\cite{Tsuei:d-wave:implications}, as it removes certain limitations of conventional circuits, \eg, allows to build RSFQ-like circuits with minimum number of bias resistors which means much lower dissipation\cite{Gerritsma:RSFQwoBias}.

In section \ref{Sec:BasicEq} we derive the version of sine-Gordon equation which describes the dynamics of the Josephson phase in LJJ with alternating $0$ and $\pi$ regions (facets). This equation describes all possible excitations, such as fluxons, semi-fluxons, plasma waves, \etc{} In section \ref{Sec:SF-shape} we obtain an explicit expression for the semi-fluxon shape and shortly discuss the properties of semi-fluxons. Section \ref{Sec:Conclusion} concludes this work and presents some ideas on future investigations of such junctions.

\section{Derivation of the basic equation}
\label{Sec:BasicEq}

\begin{figure}
  \centering
  {\includegraphics{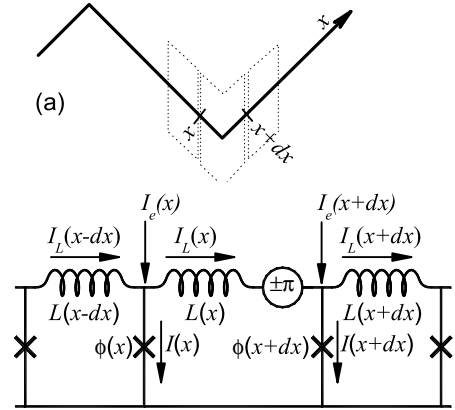}}
  \caption{
    Sketch of a fragment of a zigzag LJJ (a) and its equivalent circuit (b).
  }
  \label{Fig:circuit}
\end{figure}

We consider a zigzag LJJ as a planar 1D LJJ with a curvelinear coordinate $x$ along the zigzag, as shown in Fig.~\ref{Fig:circuit}(a). A magnetic field is applied perpendicular to the plane of the structure. Such a model can be used to describe both GB and ramp zigzag LJJs. For ramps, the direction of magnetic field can also be different, and one has to calculate the effective field following the guidelines presented in Ref.~\onlinecite{Heinsohn:RampH}. The only difference between a conventional LJJ model and our model is the presence of phase $\pi$-discontinuity points along the junction\cite{mutual-inductance}.

The Kirchhof equations for currents [Fig.~\ref{Fig:circuit}(b)] and for the Josephson phases in an elementary loop are
\begin{eqnarray}
  \phi(x+dx) -\phi(x) &=& 
  \frac{2\pi}{\Phi_0}\left[ \Phi_e - I_L(x) L(x) \right] + \Pi(x)
  , \label{Eq:Kirchhof:phases}\\*
  I_L(x) + I_e(x) &=& I_L(x+dx) + I(x)
  , \label{Eq:Kirchhof:currents}
\end{eqnarray}
where $\phi(x)$ is the Josephson phase at point $x$ of the junction, $\Phi_e(x)$ is the external magnetic flux applied to the cell, $L(x)$ is the inductance of the piece of the junction electrodes between $x$ and $x+dx$, $I_L(x)$ is the current in the electrodes, \ie, through the inductance $L(x)$, $I_e(x)$ is the externally applied bias current, and $I(x,t)$ is the current through the Josephson junction. The particular expression for $I(x,t)$ depends on the JJ model adopted and is introduced later.

The function $\Pi(x)$ can be equal to $0$ or $\pm\pi$ and shows whether there is a corner (discontinuity point) on the interval from $x$ to $x+dx$. The values $\pm\pi$ describe the direction of the jump. The function $\Pi(x)$ acts as there would be a generator of phase $\pm\pi$ inserted in the corresponding elementary loops. Imagine that we have a function $\theta(x)$ which is constant everywhere and jumps by $\pm\pi$ at each corner. This function can be written as 
\begin{equation}
  \theta(x) = \pi \sum_{k=1}^{N_c} \sigma_k \mathrm{H}(x-x_k)
  , \label{Eq:theta}
\end{equation}
where $\sigma_k=\pm 1$ defines the direction of the $k$-th jump and the sum is over all $N_c$ corners located at $x=x_k$, $\mathrm{H}(x)$ is the Heaviside step function. 

Assuming that $dx$ is smaller than the distance between the corners, the function $\Pi$ can be written as follows
\begin{equation}
  \Pi(x) = \theta(x+dx) - \theta(x)
  , \label{Eq:Pi}
\end{equation}

Assuming that the interval $dx$ is infinitesimal, we can rewrite Eqs.~(\ref{Eq:Kirchhof:phases}) and (\ref{Eq:Kirchhof:currents}) in a differential form using the following expressions: 
\begin{eqnarray}
  I &=& j(x) w\, dx
  , \label{Eq:I(x)-via-j(x)}\\
  I_e &=& j_e(x) w\, dx
  ; \label{Eq:Ie}\\
  L &=& \frac{\mu_0d'}{w}dx
  ; \label{Eq:L}\\
  \Phi_e &=& \mu_0 (\mathbf{H \cdot n}) \Lambda dx = \mu_0 H(x) \Lambda\,dx
  , \label{Eq:Phi_e}
\end{eqnarray}
where $\mu_0 d'$ is the inductance of one square of the superconducting electrodes \cite{Likharev:book}, $d'\approx2\lambda_L$ is the effective magnetic thickness of the junction\cite{Likharev:book}, ${\bf \vec{n}}$ is the unit vector normal to the plane of the junction cell as shown in Fig.~\ref{Fig:circuit}(b), $\Lambda\approx2\lambda_L$ is the effective penetration depth of the magnetic field into the junction\cite{Likharev:book}, $\lambda_L$ is the London penetration depth of the superconducting electrode, and $w$ is the width of the junction, \eg, for a GB LJJ it is equal to the film thickness. We assume that the films are spatially uniform so that $w$, $d'$ and $\Lambda$ are independent on $x$.

Substituting Eqs.~(\ref{Eq:Pi})--(\ref{Eq:Phi_e}) into Eqs.~(\ref{Eq:Kirchhof:phases}) and (\ref{Eq:Kirchhof:currents}) we rewrite them in a differential form (dividing by $dx\to0$):
\begin{eqnarray}
  \phi_x &=& \frac{2\pi}{\Phi_0}
  \left[ H\Lambda - \frac{I_L}{\mu_0d'} \right]
  + \theta_x(x)
  ; \label{Eq:Kirchhof1}\\*
  \frac{dI_L}{dx} &=& (j_e-j)w
  . \label{Eq:Kirchhof2}
\end{eqnarray}
Here and below, the subscripts $t$ and $x$ denote the partial derivatives with respect to time $t$ and coordinate $x$, respectively.

Excluding $I_L(x)$ from Eqs.~(\ref{Eq:Kirchhof1}) and (\ref{Eq:Kirchhof2}), we get the equation which describes the dynamics of the Josephson phase in the system
\begin{equation}
  (j_e-j) = \frac{1}{\mu_0d'}\left\{
    \mu_0\Lambda H_x(x) - \frac{\Phi_0}{2\pi}
    \left[ \phi_{xx} - \theta_{xx}(x) \right]
  \right\}
  , \label{Eq:Final:Phys:woModel}
\end{equation}
For the Resistively Shunted Junction model, one should substitute
\begin{equation}
  j(x) = j_c \sin(\phi) 
  + \frac{\Phi_0}{2\pi \rho} \phi_t 
  + C'\frac{\Phi_0}{2\pi}\phi_{tt}
  \label{Eq:RSJ-model}
\end{equation}
into Eq.~(\ref{Eq:Final:Phys:woModel}). Here $j_c$, $\rho$ and $C'$ are the critical current density, specific resistance and specific capacitance of the junction, respectively. After this, Eq.~(\ref{Eq:Final:Phys:woModel}) can be rewritten in a form which resembles the usual sine-Gordon equation\cite{Likharev:book}:
%
  \begin{eqnarray}
    \lambda_J^2 \phi_{xx} - \omega_p^{-2}\phi_{tt} - \sin(\phi)
    &=&\omega_c^{-1}\phi_t - \gamma(x) +
    \nonumber\\*
    &+& Q H_x(x) + \lambda_J^2\theta_{xx}(x)
    , \label{Eq:Final:Phys}
  \end{eqnarray}
%
where $\lambda_J=\sqrt{\Phi_0/(2 \pi \mu_0 j_c d')}$ is the Josephson penetration depth, $\omega_p=\sqrt{2 \pi j_c/(\Phi_0 C')}$ is the Josephson plasma frequency, $\omega_c=2 \pi j_c \rho/\Phi_0$ is the characteristic frequency, $\gamma(x)=j_e(x)/j_c$ is a normalized bias current density, and $Q=2\pi\mu_0\Lambda\lambda_J^2/\Phi_0$.

For theoretical investigation of the system we introduce standard normalized units, \ie, we normalize the coordinate to the Josephson penetration depth $\lambda_J$, and the time to the inverse plasma frequency $\omega_p^{-1}$. After such simplifications, Eq.~(\ref{Eq:Final:Phys}) can be rewritten as:
\begin{equation}
  \phi_{xx} - \phi_{tt} - \sin(\phi)
  =\alpha\phi_t - \gamma(x)
  + h_x(x) + \theta_{xx}(x)
  , \label{Eq:Final:Norm}
\end{equation}
with the damping coefficient $\alpha=\omega_p/\omega_c \equiv 1/\sqrt{\beta_c}$, and the field $h$ normalized in the usual way as $ h(x) = {2H(x)}/{H_{c1}}$, where $H_{c1}=\Phi_0/(\pi\mu_0\Lambda\lambda_J)$ is the first critical field (penetration field) for a LJJ which is, in fact, equal to the field in the center of the fluxon. From now on all quantities are given in normalized units.

In comparison with the usual perturbed sine-Gordon equation, Eq.~(\ref{Eq:Final:Norm}) contains an additional term 
\begin{equation}
  \theta_{xx}(x) = \pi\sum_k\sigma_k\delta_x(x-x_k)
  . \label{Eq:theta1}
\end{equation}
which describes the corners with $\pm\pi$ phase jumps. 

To simplify the analysis it is convenient to present the phase $\phi$ as a sum of two components: the magnetic one $\mu(x)$ and the order-parameter related one $\theta(x)$ (\ref{Eq:theta}), \ie,
\begin{equation}
  \phi(x,t) = \mu(x,t) + \theta(x)
  . \label{Eq:phi=mu+theta}
\end{equation}
In this case we can get rid of $\delta$-functions in the Eq.~(\ref{Eq:Final:Norm}) and rewrite it only for the ``magnetic'' component $\mu$:
\begin{equation}
  \mu_{xx} - \mu_{tt} - \sin(\mu)\underbrace{\cos(\theta)}_{\pm1}
  =\alpha\mu_t - \gamma(x)
  + h_x(x)
  . \label{Eq:Final:mu:Norm}
\end{equation}
It is rather interesting that this is just the usual perturbed sine-Gordon equation, but the sign of $\sin(\mu)$ changes from facet to facet. This means that every second facet can be considered as having negative critical current $-1$ (in normalized units) instead of $+1$. Note, that this is only valid for a current--phase relation with odd harmonics. In the general case odd harmonics change the sign, and even harmonics do not. This applies to both sine and cosine harmonics in the Fourier representation of the current--phase relation.

It may be easier, especially for numerical investigations, to use Eq.~(\ref{Eq:Final:mu:Norm}) instead of Eq.~(\ref{Eq:Final:Norm}). Eq.~(\ref{Eq:Final:mu:Norm}) can be solved separately on each interval between corners, and all solutions should be joined at $x=x_k$. On the other hand, solving Eq.~(\ref{Eq:Final:mu:Norm}) implies dealing with $\delta$-functions, which may be rather cumbersome.

\section{Semi-fluxon}
\label{Sec:SF-shape}

As was found experimentally \cite{Hilgenkamp-Tsuei}, the presence of $\pi$ discontinuities of the phase, may result in the formation of semi-fluxons pinned at the corners of the zigzag. Let us consider an infinitely long LJJ with a single corner at $x=0$ and derive an analytical expression which describes the shape of such a semi-fluxon. We start from the static version of Eq.~(\ref{Eq:Final:mu:Norm}) without perturbation terms
\begin{equation}
  \mu_{xx}^{\mp}=\pm\sin\mu
  , \label{Eq:sG0}
\end{equation}
where we have assumed that $\theta(x) = -\pi\mathrm{H}(x)$, \ie, phase jumps from $0$ to $-\pi$ when we pass the corner at $x=0$. $\mu^-(x)$ refers to the left half of LJJ ($x<0$), while $\mu^+(x)$ refers to the right half ($x<0$). The semi-fluxon is generated to compensate a phase jump at $x=0$, and, far from the corner, the LJJ should not ``know'' about the jump. Therefore, we search for a solution of Eq.~(\ref{Eq:sG0}) which has the following boundary conditions at infinity
\begin{equation}
  \phi(\pm\infty)=0
  ; \text{ and }\quad 
  \phi_x(\pm\infty)=0
  . \label{Eq:BC:phi}
\end{equation}
The same conditions for $\mu(x)$ are
\begin{eqnarray}
  \mu(-\infty)=0; \quad \mu(+\infty)=\pi
  ; \label{Eq:BC:mu}\\*
  \mu_x(\pm\infty)=0
  . \label{Eq:BC:mu'}
\end{eqnarray}

We multiply both sides of Eq.~(\ref{Eq:sG0}) by $2\mu_x^\mp$ and rewrite it in the form
\begin{equation}
  \left[ \left( \mu_x^\mp \right)^2 \right]_x = \mp 2 (\cos\mu^\mp)_x
  . \label{Eq:sG1}
\end{equation}
After integration we get
\begin{equation}
  \left( \mu_x^\mp \right)^2 = \mp 2 \cos\mu_\mp + C
  . \label{Eq:sG_IntC}
\end{equation}
The integration constant $C$ can be determined from the conditions (\ref{Eq:BC:mu}). Taking the limit of Eq.~(\ref{Eq:sG_IntC}) at $x\to\pm\infty$ we see that Eq.~(\ref{Eq:sG_IntC}) holds only provided $C=2$. Thus
\begin{subequations}
  \begin{eqnarray}
    \left( \mu_x^- \right)^2 = 2 (1-\cos\mu^-) = 4 \sin^2\frac{\mu^-}{2}
    ; \label{Eq:sG_Int-}\\*
    \left( \mu_x^+ \right)^2 = 2 (1+\cos\mu^+) = 4 \cos^2\frac{\mu^+}{2}
    . \label{Eq:sG_Int+}
  \end{eqnarray}
  \label{Eq:sG_Int}
\end{subequations}
Let us introduce a new variable $\psi^\pm=\mu^\pm/2$ and take the square root of both parts of Eq.~(\ref{Eq:sG_Int}). We suppose that $\mu^{\pm}_x\ge0$ for all $x$, \ie, $\mu$ grows from $0$ to $\pi$ always with non-negative derivative. This can be checked later when we will get a solution. Therefore we keep only the plus sign in front of the square root (the minus sign corresponds to a negative semi-fluxon), so we get
\begin{subequations}
  \begin{eqnarray}
    \psi^{-}_x = \sin\psi^-
    ; \label{Eq:sG_IntPsi-}\\*
    \psi^+_x = \cos\psi^+
    . \label{Eq:sG_IntPsi+}
  \end{eqnarray}
  \label{Eq:sG_IntPsi}
\end{subequations}
Integrating this equation yields
\begin{subequations}
  \begin{eqnarray}
    x + x^-_* &=& 
    \int\frac{d\psi^-}{\sin\psi^-} =
    \ln\tan\frac{\psi^-}{2}
    . \label{Eq:x(phi)-}\\*
    x + x^+_* &=& 
    \int\frac{d\psi^+}{\cos\psi^+} = 
    \ln\frac{1+\sin\psi^+}{\cos\psi^+} 
    . \label{Eq:x(phi)+}
  \end{eqnarray}
  \label{Eq:x(phi)}
\end{subequations}
Using the condition $\psi(0)=\mu(0)/2=\pi/4$, we can determine the value of the integration constant $x_*$:
\begin{subequations}
  \begin{eqnarray}
    x^-_* &=& 
    \ln\tan\frac{\pi}{8} = \ln\left( \sqrt{2}-1 \right) = \ln{\cal G}
    . \label{Eq:x_0-}\\* 
    x^+_* &=& 
    \ln\frac{1+\sin\pi/4}{\cos(\pi/4)} = 
    \ln\left( \sqrt{2}+1 \right) = \ln\frac{1}{\cal G}
    , \label{Eq:x_0+}
  \end{eqnarray}
  \label{Eq:x_0}
\end{subequations}
where ${\cal G}=\tan(\pi/8)=\sqrt{2}-1\approx0.404$.

Solving the Eq.~(\ref{Eq:x(phi)}) for $\psi=\mu/2$ and using (\ref{Eq:x_0}) we get
\begin{subequations}
  \begin{eqnarray}
    \mu^-(x) &=& 4 \arctan\left( {\cal G} e^{x} \right) 
    . \label{Eq:semi-fluxon-}\\* 
    \mu^+(x) &=& 4 \arctan\frac{1-{\cal G}e^{-x}}{1+{\cal G}e^{-x}}
    \nonumber\\*
    &=& \pi - 4 \arctan\left( {\cal G}e^{-x} \right)
    . \label{Eq:semi-fluxon+}
  \end{eqnarray}
  \label{Eq:semi-fluxon}
\end{subequations}

The final expression for the semi-fluxon shape in terms of the total phase $\phi(x)$ can be written in a more compact form as
\begin{equation}
  \phi(x) 
  = - 4\mathrm{sign}(x) \arctan\left( {\cal G} e^{-|x|} \right) 
  , \label{Eq:SF}
\end{equation}

If we want to calculate the magnetic field we should not forget that $\phi(x)$ consists of two components: (1) $\theta(x)$ phase jumps at the corners, and (2) the magnetic component $\mu(x)$ which describes the smooth variation of the phase. It is the derivative of $\mu(x)$ that is equal to the magnetic field at a given point of the junction. From Eq.~(\ref{Eq:semi-fluxon}) the field $\mu_x(x)$ is given by the expression
\begin{equation}
  \mu_x(x) = \frac{2}{\cosh(|x|-\ln{\cal G})}
  \quad . \label{Eq:mu_x}
\end{equation}
The field in the center of the semi-fluxon is
\begin{equation}
  \mu_x(0) = \frac{2}{\cosh\ln{\cal G}}
  = \frac{4}{{\cal G}+\frac{1}{\cal G}}   
  = \sqrt{2}  
  , \label{Eq:mu_x(0)}
\end{equation}
and should be compared with the field in the center of a fluxon, which is equal to $2$.

The supercurrent density can be calculated as
\begin{equation}
  \sin(\phi)=\mu_{xx} = -2\mathrm{sign}(x)
  \frac{\sinh(|x|-\ln{\cal G})}{\cosh^2(|x|-\ln{\cal G})}
  . \label{Eq:sin(phi)}
\end{equation}
\begin{figure}
  \centering
  {\includegraphics{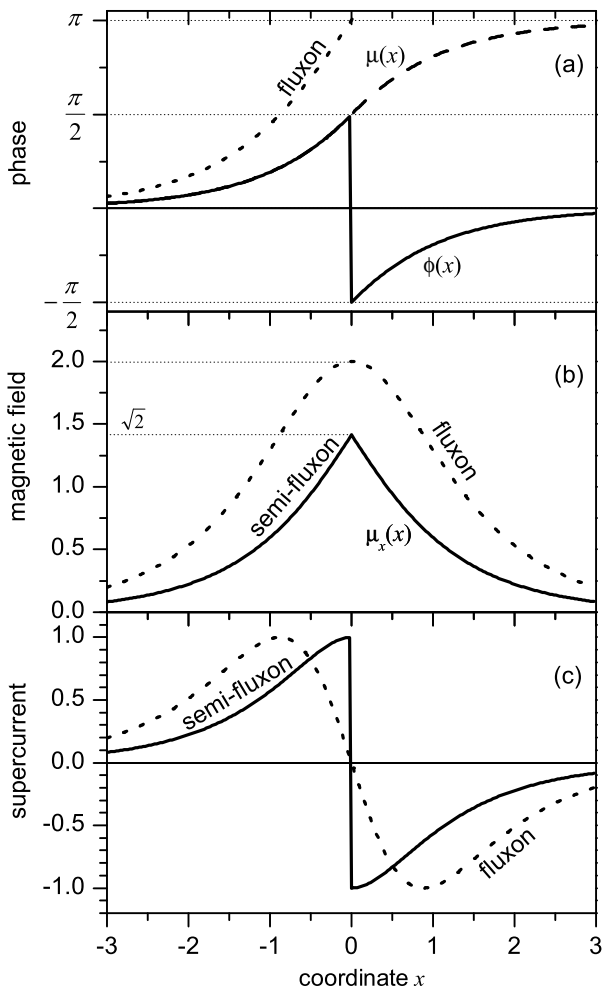}}
  \caption{
    Comparison of fluxon and semi-fluxon shapes.
    (a) The behavior of total phase $\phi(x)$ and of magnetic component $\mu(x)$ only. 
    (b) Magnetic field profile $\mu_x(x)$.
    (c) Supercurrent profile $\sin(\phi)=\mu_{xx}(x)$.
  }
  \label{Fig:SF}
\end{figure}

The functions $\phi(x)$, $\mu(x)$, $\mu_x(x)$ (magnetic field), and $\sin(\phi)$ (supercurrent) are shown in Fig.~\ref{Fig:SF}. The very difference between a fluxon and a semi-fluxon is that (a) the fluxon carries one quantum of magnetic flux, while the semi-fluxon carries only half of the flux quantum (therefore the name), and (b) the semi-fluxon has a sharp maximum which looks like a cusp. It would be very interesting to compare the shape of a SF obtained by scanning SQUID microscopy 
\cite{Hilgenkamp-Tsuei} with the shape given by 
Eq.~(\ref{Eq:mu_x}). 

The Eq.~(\ref{Eq:SF}) describes a positive semi-fluxon (PSF), {\ie} the one containing $+\Phi_0/2$. To describe a negative semi-fluxon (NSF), one just have to alter the sign in front of Eq.~(\ref{Eq:SF}) or to change $x\to-x$ in Eq.~(\ref{Eq:semi-fluxon}). In the same time, one should keep in mind that the sign of $\theta$ [of $\sigma_k$ in Eq.~(\ref{Eq:theta})] is in no way related to the polarity of the SF. One can as well construct the PSF which sits at the point where $\theta$ jumps up from $0$ to $+\pi$. In this case the total phase twist will be equal to $2\pi$, but physically the situation will not change. Shortly, it is only the sign of $\mu$, but not $\theta$ [$\sigma_k$ in Eq.~(\ref{Eq:theta})] which defines the polarity of the SF.

SFs are very similar to fluxons when they interact with each other: semi-fluxons and (semi-)fluxons of the same polarity repel each other, while the ones of opposite polarity attract themselves. This can be easily shown writing the potential energy as a function of the distance between them.

\section{Conclusions}
\label{Sec:Conclusion}

We have derived the perturbed sine-Gordon equation which describes the dynamics of the Josephson phase in a LJJ containing phase $\pi$-discontinuities, which correspond to the corners of the Nb-YBCO zigzag LJJ. Using the derived basic Eq.~(\ref{Eq:Final:Norm}) we have obtained the shape of a semi-fluxon --- the new type of object which appears due to the phase jumps. Our results allow to investigate the interaction between semi-fluxons as well as between fluxons and semi-fluxons. This is also a starting point for the numerical simulation of various aspects of fluxon and semi-fluxons dynamics. 

An interesting consequence of Eq.~(\ref{Eq:Final:Norm}) is that the terms $\gamma(x)$ and $\theta_{xx}(x)$ play a similar role in the equation. This means that one can substitute one by the other. For example, one may wish to create an artificial $\pi$-discontinuity point using only superconductors with $s$-wave order parameter symmetry with an injector and receptor of current of the size $\Delta x$ separated by a minimum distance. Passing the current equal to $4\pi/\Delta x^2$ from the injector to the receptor, one emulates the effect of $\theta_{xx}$. Of course, $\Delta x$ must be much smaller than any characteristic length, \eg, $\Delta x \ll \lambda_J$. Although this emulation may look not ideal, one should keep in mind that in a real zigzag junction, the size of the corner is also finite and is defined by the lithographic accuracy $\sim 1\mathrm{\mu m}$.

\begin{acknowledgments}
  We would like to thank H.~Hilgenkamp, H.-J. Smilde and C.~C.~Tsuei for stimulating discussions.
  
\end{acknowledgments}


\begin{thebibliography}{99}
\bibitem{mutual-inductance}
  For simplicity we neglect mutual inductances which arise due to corners of the zigzag and give a larger contribution than in the case of a usual linear LJJ.
\end{thebibliography}
\end{document}